\documentclass[pra, aps, showpacs, superscriptaddress, twocolumn]{revtex4-2}
\usepackage{graphicx,amsmath,amssymb,color}
\usepackage{physics}
\usepackage{graphicx}
\usepackage{epstopdf}
\usepackage{float}
\usepackage{xcolor}
\usepackage[colorlinks, citecolor={blue!50!black}, urlcolor={blue!50!black}, linkcolor={red!50!black}]{hyperref}
\newcommand{\eq}[1]{\begin{align}#1\end{align}}
\begin{document}

\title{Quantum Random Number Generator (QRNG): Theoretical and Experimental Investigations}

\author{Zeshan Haider\textsuperscript{*}}
\affiliation {National Institute of Lasers and Optronics College, Pakistan Institute of Engineering and Applied Sciences, Nilore, Islamabad $45650$, Pakistan.}
\email{Corresponding author: shani12441@gmail.com}
\author{Muhammad Haroon Saeed}
\affiliation {National Institute of Lasers and Optronics College, Pakistan Institute of Engineering and Applied Sciences, Nilore, Islamabad $45650$, Pakistan.}
\author{Muhammad Ehsan-ul-Haq Zaheer}
\affiliation{National Institute of Lasers and Optronics College, Pakistan Institute of Engineering and Applied Sciences, Nilore, Islamabad $45650$, Pakistan.}
\author{Zeeshan Ahmed Alvi}
\affiliation {National Institute of Lasers and Optronics College, Pakistan Institute of Engineering and Applied Sciences, Nilore, Islamabad $45650$, Pakistan.}
\author{Muhammad Ilyas}
\affiliation {National Institute of Lasers and Optronics College, Pakistan Institute of Engineering and Applied Sciences, Nilore, Islamabad $45650$, Pakistan.}
\author{Tahira Nasreen}
\affiliation {National Institute of Lasers and Optronics College, Pakistan Institute of Engineering and Applied Sciences, Nilore, Islamabad $45650$, Pakistan.}
\author{Muhammad Imran}
\affiliation {National Institute of Lasers and Optronics College, Pakistan Institute of Engineering and Applied Sciences, Nilore, Islamabad $45650$, Pakistan.}
\author{Rameez Ul Islam}
\affiliation {National Institute of Lasers and Optronics College, Pakistan Institute of Engineering and Applied Sciences, Nilore, Islamabad $45650$, Pakistan.}
\author{Manzoor Ikram}
\affiliation {National Institute of Lasers and Optronics College, Pakistan Institute of Engineering and Applied Sciences, Nilore, Islamabad $45650$, Pakistan.}

\date{\today}

\begin{abstract}
Quantum Random Number Generators (QRNGs) emerged as a promising solution for generating truly random numbers. In the present article, we give an overview of QRNGs highlighting the merits and demerits of various strategies briefly. Then opting for the best-case scenario, we present the in-depth experimental explorations for building and characterizing QRNG using the homodyne detection technique to measure the quadrature amplitude of quantum vacuum fluctuations. Since entropy assessment plays a fundamental role in authenticating the true randomness, a comprehensive description of entropy and how it evaluates the quality of randomness of the source is illustrated. Our experimental setup, apart from the hardware, includes a diverse set of testing techniques including NIST statistical/entropy suites, Dieharder tests battery, and autocorrelation coefficient to verify the randomness and statistical properties of the generated random numbers. We believe that our experimental investigations provide a valuable resource for building QRNGs for a wide range of applications. 
\end{abstract}

\maketitle
\section{introduction}
The need for random numbers has been a long-standing requirement in many fields, however, generating truly random numbers has been a challenging task due to the deterministic nature of classical backing such a number generation. Prior to the advent of quantum technologies, abstract one-way mathematical functions were employed to generate random numbers which were hard enough to guess though sequential guessing is quite possible through the utilization of quantum computing algorithms ~\cite{inproceedings,doi:10.1080/01621459.1949.10483310,schneier2015applied}. Quantum mechanics has opened up new avenues for generating true randomness with its fundamental properties of uncertainty and inherent indeterminism. Quantum Random Number Generators (QRNGs) use this property to furnish random numbers that cannot be predicted or replicated and, are widely used in many applications that require high-quality randomness\cite{Schmidt1970QuantumMechanicalRG,RevModPhys.89.015004,ma2016quantum}. The various techniques of QRNG has been introduced and demonstrated so far including the detection of arrival time of photons, path-distinguished of single photons passed through Beam splitter, distribution of spontaneously emitted photons \cite{Michler1998AnET} and the measurement of quantum vacuum fluctuations \cite{rarity1994quantum,stefanov2000optical,jennewein2000fast,ma2005random,dixon2008gigahertz,wayne2010low,furst2010high,qi2010high,gabriel2010generator,zhang2016note,jofre2011true,bustard2011quantum,jian2011two,marandi2012all,xu2012ultrafast,nie2014practical,yan2014multi,nie2015generation}. Apart from this, QRNGs have also been experimentally demonstrated based on the quantum nonlocality of entangled photons pairs \cite{pironio2010random,lunghi2015self}, phase noise of lasers\cite{qi2010high,guo2010truly,nie2015generation,xu2012ultrafast,jofre2011true,abellan2015generation,abellan2014ultra,yuan2014robust}, photon number distribution\cite{wei2009bias,furst2010high,ren2011quantum,yan2014multi,applegate2015efficient}, and other methods \cite{liu2018secret,wang2006scheme,zhou2017quantum}. The effort to expand the Poissonian nature of light to super-Poissonian through the scattering process has also been made to suppress photonic correlations leading to the broadening of the random numbers spectrum\cite{dandasi2019optical,haider2023multiphoton}. In applications like cryptography \cite{saeed2022implementation}, random numbers from untrusted sources may result in serious safety issues and it is quite possible for the hackers to crack the encryption system if they get access to the decodable random numbers\cite{bouda2012weak}. The development of cryptographic technologies like quantum key distribution\cite{bouda2012weak} demands certified, real-time, and high-speed random number generation, which has accelerated this field of research unarguably. This intrinsic randomness, derived from the probabilistic nature of quantum phenomena, is indispensable for propelling advancements in quantum computation and fortifying the security foundations of quantum communication \cite{liu2023source,nie2015generation,xie2022breaking,lucamarini2018overcoming,gu2022experimental,chen2021twin,chan2021stable,yin2023experimental}.

In this paper, we present a comprehensive overview of the basic concepts involved in defining and characterization of random numbers with due emphasis over entropic investigations, a major theme while dealing with true random numbers. Next, we furnish the detailed engineering schematics with a due evaluation of the generated numbers through standard tests incorporating entropy assessment as the major ingredient. In this context, our experiment is based on homodyne measurement of the amplitude quadrature of quantum vacuum state through indigenously fabricated $10$MHz Balanced Homodyne Detector (BHD), laser operating at $852$nm wavelength along with the rest of essential optical components. The post-processing performed over the raw data includes the digitization of analog noise, the implementation of an extractor algorithm to minimize the inevitable correlations due to classical noise sources, and the testing of the random bits through potential statistical test suites.
\subsection{Types and Methods of Random Number Generators (RNGs)}
A lot of efforts have been devoted to finding the reliable ways to harness the nondeterministic and true random numbers in the last few decades \cite{1890Natur..42...13G,von195113}. Such efforts owe to the extensive applications of the randomness in software developments \cite{Ghosh2021RandomnessIT}, quantum algorithms \cite{Melnikov2021BenchmarkingQR}, quantum cryptography \cite{Pironio2021LoopholefreeBF}, lottery protocols \cite{Iqbal2021SecureLP}, Monte-Carlo Simulations \cite{Salloum2021RandomNP} among others. In this context, the real-time speed of the random bit stream is one of the promising challenges as most of the applications require millions of random inputs (triggers) per second. A good RNG is, for instance, characterized through its speed (random-bits/Sec) along with its versatility and authenticity level. The RNGs are, in general, categorized in two classes: Pseudo RNGs and Physical RNGs. Pseudo RNGs rely on the abstract mathematical algorithms and do not require any physical seed to initiate. These algorithms include  congruential generator \cite{schneier2007applied,trappe2006introduction,knuth2011art}, Mersenne Twister \cite{matsumoto1998mersenne} and the Blum-Blum-Shub(BBS) generator \cite{schneier2007applied,trappe2006introduction,blum1983comparison}. The output i.e., stream of 0s and 1s of such RNGs appears to be truly random, having negligible correlations and fulfills various randomness testing criteria. However, these mother algorithms for pseudo RNGs are deterministic and have become more compromised with the recent advancement of the quantum computing empowered with extremely efficient search algorithms. This class of random bits is thus not applicable for cryptographic purposes.

The Physical RNGs, on the other hand, are associated with the physical source or seed to generate true random numbers. They are based on hardware and require a physical process, for instance, the measurement of the noisy system  to generate random bits. In physical RNGs, it is essential to consider an appropriate physical process as an entropy source and such a source can be either based on a process described by classical physics or by the quantum theory.  The various noisy systems have been employed to generate random bits ranging from spontaneous radioactive decays and  atmospheric noise to  human behavior-related events. The electronic noise of the complex circuits e.g. chona's circuit can also be used as a potential source of randomness. The randomness from keystroke's timings, physical movements of the mouse as well as environmental noises have already been harnessed in the UNIX-like operating system \cite{Gaurav2021AnAO}. The question still important and relevant is that can macroscopic physical processes based on classical mechanics be truly unpredictable? The fair coin tossing, a macroscopic process, becomes partially (or completely) biased if one can able to predict the coupling of coin with the environment. The physical RNG becomes perfectly credible at the microscopic level subjected to the quantum mechanical  descriptions. The inherent nature of unpredictability of a quantum system can thus be exploited to obtain the true random numbers. In the following, a brief comparison between Quantum and classical description-based random numbers is given.
\subsection{Random Number Generators in the Realm of Classical and Quantum Theory}\label{sec:TPB}
Classical physics is the collective knowledge gathered and formulated by the physicists over hundreds of years, which describes macroscopic systems like falling coins under a toss scenario. Quantum physics is a set of theories elaborated by physicists during the early decades of the 20th century and explains the dynamics of the microscopic systems like atoms or elementary particles. This description though mathematically quite simple is highly counterintuitive, nonlocal and inherently random at the level of the single or few quantum entities. This randomness emanates from the concept like superposition, entanglement and quantum vacuum are completely mind-boggling and philosophically very hard to grasp.

Macroscopic processes described by classical physics are often utilized to generate random numbers. The most frequent and common random number generators i.e., coin and dice tossing – indeed belongs to this class. The dynamics of a system explained by classical physics can be predicted, under the supposition that the initial conditions are known. Now for a coin, a physicist knowing precisely its weight, its initial position, the force applied to it by the hand, the speed of the wind, as well as all the other relevant parameters, should in principle be able to predict the outcome of the throw. Why this prediction is not possible in practice? Coin tossing is a chaotic process. Chaos is a type of behavior observed in systems whose evolution exhibits extreme sensitivity to initial conditions. Coin tossing is not the only physical system with chaotic evolution. There are a lot of  others e.g. turbulences in a flow (turbulences in a lava lamp have been used to generate random numbers) \cite{Bai2021LavaLR} or meteorological phenomena with butterfly type effects are good examples of chaotic systems \cite {Luo2021QuantumNR}. The evolution and behavior of these systems are highly sensitive to initial conditions. In spite of its popularity, coin tossing is clearly not very practical when many random events need to be known to reach a stable quality of randomness.

Contrary to classical physics, quantum physics is fundamentally and inherently random and nature offers no further explanation for such randomness. Its intrinsic randomness has been confirmed over and over again by theoretical and experimental research conducted since the dawn of the $20^{th}$ century. When designing a random number generator, it is a natural choice to take advantage of this intrinsic randomness and to resort to the use of a quantum process as a source of randomness.  The first quantum random number generators were based on the temporal observation of the radioactive decay of some specific elements. Here quantum tunneling is the reason beyond nuclear decay and any radioactive particles e.g. $\alpha$, or $\beta$-particle is always in the superposition of reflected and transmitted through the nuclear boundary as it oscillates within the nuclear surface. Although radioactive decay samples produce random numbers of excellent quality but these generators are quite bulky and the use of radioactive materials may cause general health concerns including cancer.
\maketitle
\section{Experimental Setups for QRNGs}\label{sec:TPB}
\begin{figure}[htb]
	\includegraphics[scale=0.32]{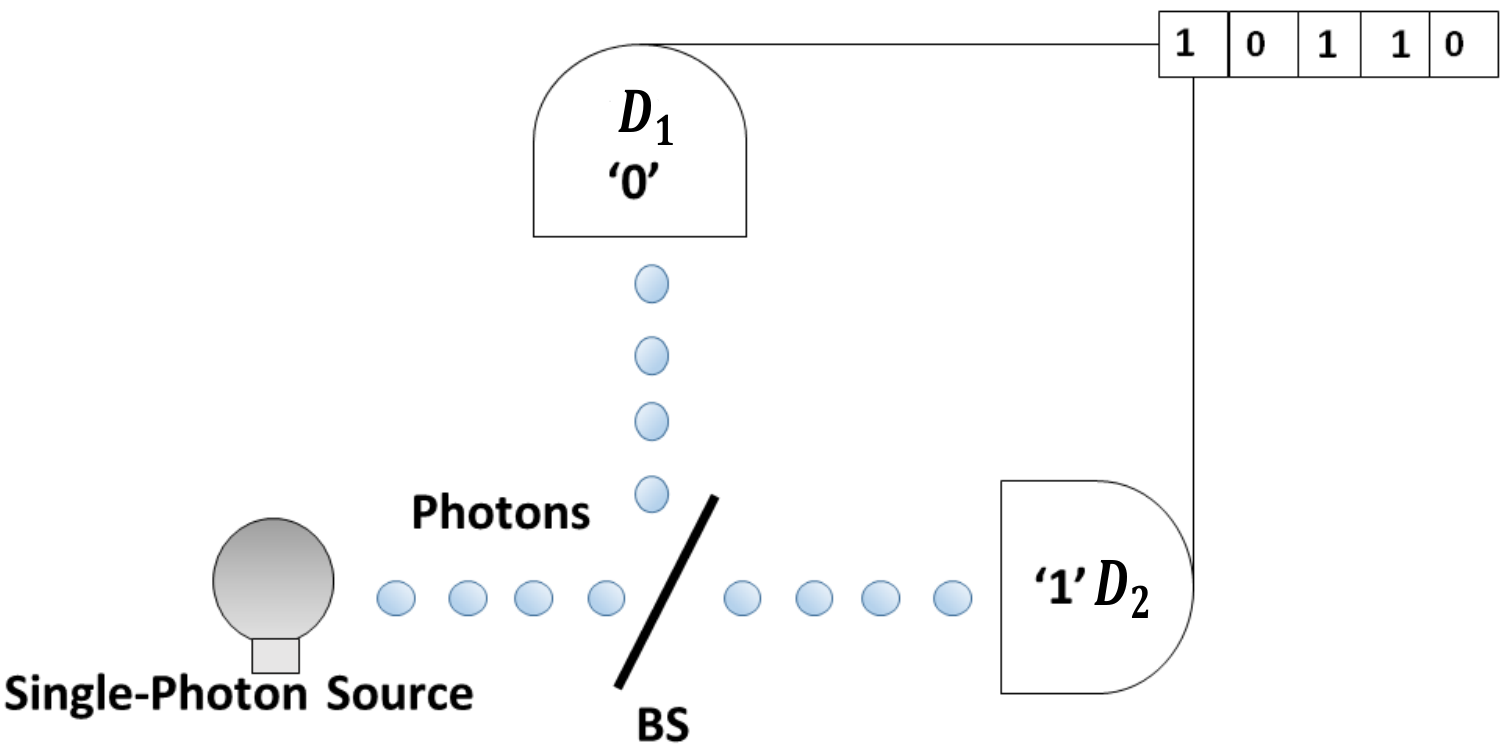}
	\caption{QRNG based on single photons and a 50:50 Beam Splitter (BS) is shown. A single photon emitted from the source gets transmitted or reflected from BS in a purely random way and clicks one of the detectors D1 or D2 labeled as a bit '$0$' or '$1$', respectively.  }\label{fig:spd}
\end{figure}
The very first idea of QRNG was presented by Anton Zeilinger\cite{jennewein2000fast} and is explained as follows: A single photon source generates the train of single photons being sent through a beam splitter, which has two output ports as shown in Fig. \ref{fig:spd}. A single photon impinging a beam splitter is both transmitted and reflected simultaneously and forms a symmetric superposition of spatial modes. The reflected or transmitted probability of the photon is thus $50:50$, so there is an equal chance of a photon being detected at either output port.
The single photon detectors $D_1$ and $D_2$ are placed at each output port to detect whether the photon is being reflected or transmitted. If the photon is detected at one output port ($D_1$), this corresponds to a value of 0, and if it is detected at the other output port ($D_2$), it corresponds to a value of $1$. To generate a random binary sequence, this process is repeated many times, with each photon representing a single bit. The resulting sequence of $0$s and $1$s is a truly random sequence that can be used for various applications. The potential limitation of such a QRNG is the low generation rate due to low detection efficiency and dark counts of the single-photon detectors and can be enhanced to some extent by using entangled photons \cite{kumar2019quantum}. Another potential limitation is the susceptibility of this method to certain types of attacks, such as side-channel attacks that exploit weaknesses in the detection equipment. For example, a recent study by Bao et al. ($2021$) showed that QRNG systems based on single photons and beam splitters may be vulnerable to certain types of photon-number-splitting attacks, which could compromise the randomness of the generated numbers \cite{bao2021vulnerability}. There are also practical limitations related to the cost and complexity of implementing a QRNG system based on single photons and beam splitters. These systems typically require specialized equipment such as single-photon detectors and precise optical components, which can be expensive and difficult to maintain \cite{huang2019fully}. Further, efficient single-photon sources often require extensive cooling and a dark room environment, etc.
\begin{figure}[htb]
	\includegraphics[scale=0.29]{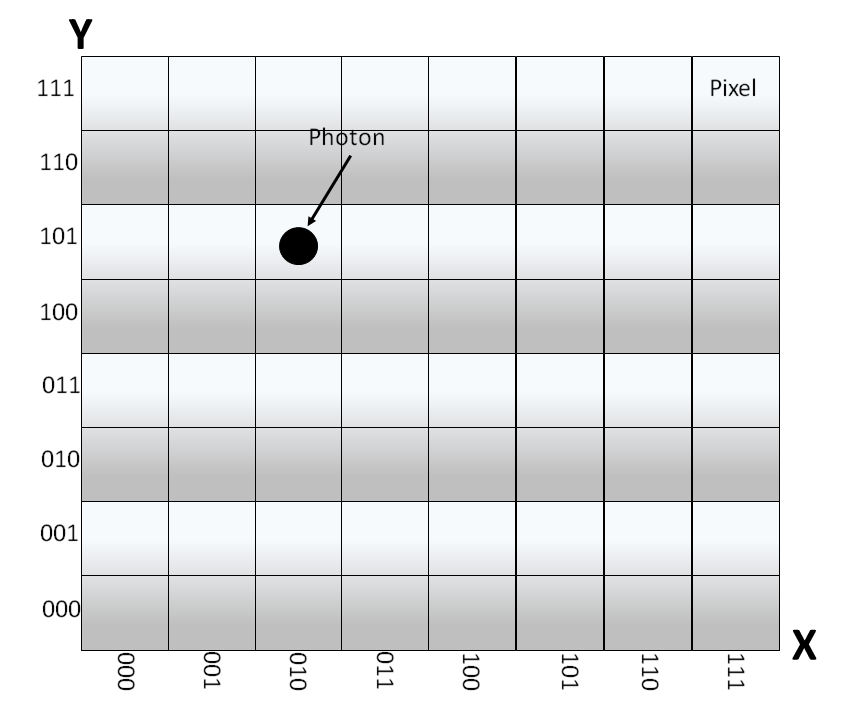}
	\caption{A facial view of planner single photon detector whose each coordinate (X and Y) is encoded with three bits. Therefore, $64$ distinct pixels make the detector to encode the position of striking photons.)}\label{fig:spatial}
\end{figure}

QRNG based on the spatial distribution of photons is one of the promising and high-speed random bit generators \cite{Li2021QuantumRN}. It relies on the single photon detector having the ability for continuous measurement of the positions of individual photons. A beam splitter randomly directs photons to the reflected or transmitted paths (similar to the above technique of single-photon QRNG). The filters that the photons pass through further add randomness to the system, as the filters are randomly oriented, implying that the photons have an equal chance of being polarized in any direction. After passing through the filters, the photons are directed onto a detector array screen, which is typically divided into many encoded pixels shown in Fig. \ref{fig:spatial}. When a photon hits a pixel, it generates an electrical signal that is recorded by the detector. The location of the photon's impact on the detector screen is used to generate a sequence of random numbers.
The use of encoded pixels allows for increased efficiency and security in the QRNG. Each pixel is assigned a unique code, and the electrical signals generated by the photons hitting the pixels are also encoded with this information. The photon shown in Fig. \ref{fig:spatial} is, for example, identified by its $3$ bit coordinates [($010,101$)], and in that planer detector, there are $2^3\times2^3=64$ choices (pixels) for a photon to be detected. This allows for easy identification of which pixel was hit by which photon, and ensures that any attempts to intercept or manipulate the signal would be easily detected.
Additionally, the use of a detector with many pixels allows for increased speed and resolution in the QRNG. By using a large number of pixels, it is possible to detect many photons simultaneously, leading to a faster generation of the random numbers. However, this class of QRNGs is device dependent and  can be affected by imperfections in the measurement equipment, precision of encoded pixels of detectors, and environmental noise \cite{marangon2017practical}.

\begin{figure}[htb]
	\includegraphics[scale=0.275]{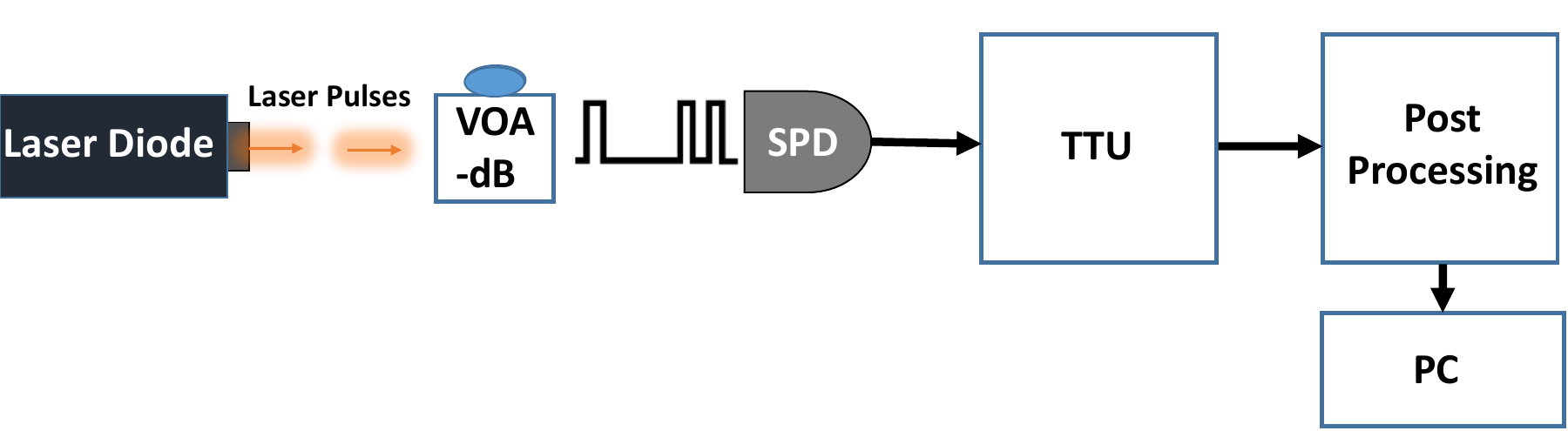}
	\caption{The schematic diagram for QRNG based on photon-arrival-time.}\label{fig:arrive}
\end{figure}

The schematic diagram for a Quantum Random Number Generator (QRNG) based on Photon Arrival Time is depicted in Fig. \ref{fig:arrive}. The first instance of such a system was demonstrated in the research by Stipvcevic et al. \cite{stipvcevic2007quantum}, where the arrival times of successive photons were compared, and bit values were assigned based on the temporal lapse with the priority given to the one that was longer. The QRNG setup begins with the generation of photons from a source, specifically, a Laser Diode (LD) equipped with an electronic pulsar that facilitates the emission of optical pulses. The optical output is significantly attenuated using a Variable Optical Attenuator (VOA) until the light level reaches optimally to that of single photons. These single photons are then detected by a Single Photon Detector (SPD). The SPD's output consists of a sequence of digital pulses, with intervals between detections determined by the photon statistics of the light source. A Time Tagging Unit (TTU) registers these detector pulses and produces multi-bit values corresponding to the time intervals between detections. These bit-values represent the random timing information, but their degree of randomness is somewhat reduced due to their respective probability distribution curve. To address this, the data undergoes whitening to render the values suitable for random number generation. The processed bit-values can subsequently be transmitted to a computer for further analysis or randomness testing.

In the context of the security of QRNG, Measurement-Device-Independent (MDI) schemes have also been employed recently \cite{liu2018device, liu2018high, pironio2010random}. We briefly present the MDI protocols for QRNG, namely (I). Entangled Photon Pair Scheme, (II). Measurement-Independence by Random Measurement Basis.

Entangled Photon Pair Scheme:
In this scheme, two entangled photons are generated, typically through Spontaneous Parametric Down Conversion (SPDC). One photon (photon-A) is sent to the user i.e., Alice, and the other photon (photon-B) is sent to the user commonly referred to as Bob. Both Alice and Bob perform random basis measurements on their respective photons. The measurement results, which can be either 0 or 1, are communicated to a third party, Charlie, who then performs an XOR operation on the results. The XORed outcome is used as a random bit since it is independent of the measurement devices used by Alice and Bob \cite{acin2007device}. Any attempt to tamper with the photons or the measurements will affect the correlations between the entangled photons, rendering it detectable by Charlie.

Measurement-Independence by Random Measurement Basis:
This scheme relies on the idea of using a random measurement basis that is independent of the quantum states being measured \cite{marcikic2006free}. The setup includes a single-photon source that generates individual photons. The photons are then sent to different measurement stations. Each measurement station is equipped with multiple detectors sensitive to different measurement bases, e.g., rectilinear (H, V) or diagonal (D, A) polarization. The measurement basis used at each station is randomly chosen for every photon. Once the measurements are performed, the outcomes are duly recorded. To generate random bits, the measurement basis information is revealed after the measurements, and a post-processing algorithm processes the measurement results based on the random basis information. Thus, by ensuring that the measurement bases are random and independent of the quantum state, this scheme achieves measurement-device independence.

\begin{figure}[htb]
	\includegraphics[scale=0.34]{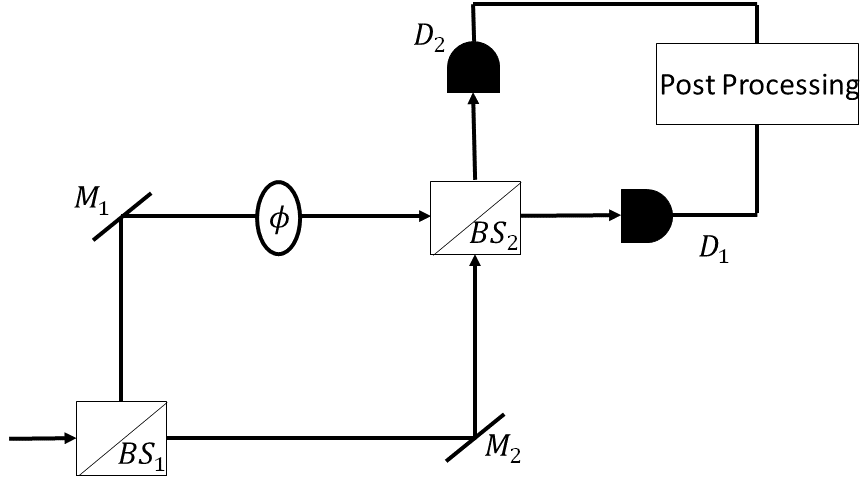}
	\caption{The illustrative diagram for the engineering schematics of the QRNG by exploiting the phase noise of a coherent photonic state. An unbalanced Mach–Zehnder Interferometer is shown followed by two detectors $D_1$ and $D_2$.  }\label{fig:phase}
\end{figure}

Another way of building QRNG is the fluctuations in the coherent quantum state (of laser) when realized in phase space. They can be a potential source of quantum randomness due to Heisenberg's uncertainty principle. These fluctuations occur in amplitude as well as phase quadratures of a state being represented in phase space and are experimentally extractable. The traditional phase noise schemes utilize a self-delayed interference structure \cite{lei20208,qi2010high} based on the unbalanced Mach‒Zehnder interferometer (MZI) as shown in Fig. \ref{fig:phase}. In an unbalanced Mach-Zehnder Interferometer, one arm (upper arm in Fig. \ref{fig:phase}) is longer than the other and their path difference, consequently, acts as a phase shifter ($\phi$) responsible for the quantum interference. The coherent laser beam at $BS_1$ splits into two beams and these beams are allowed to interfere at $BS_2$ with a mutual phase  shift $\phi$.  The random fluctuations in the phase quadrature of the coherent state of light appear in the amplitude at the output which can be detected through the detectors $D_1$ and $D_2$. This technique transforms the quantum phase noise into amplitude fluctuations that are experimentally realizable. However, a very fine and stable tuning of the temporal delay between both the paths of the interferometer is quite difficult to tackle. Furthermore, the bandwidth of phase noise in the laser necessitates a phase shifter for the unbalanced Mach-Zehnder Interferometer (MZI), which typically requires a lengthy delay-line of several meters to effectively reduce the auto-correlation coefficient of raw data. In real-world implementations, compacting the space required for this delay-line is a significant challenge unless novel approaches are employed that eliminate the need for a highly imbalanced MZI \cite{app10072431}. Such type of QRNGs typically requires calibration procedures to ensure accurate measurement of the phase noise \cite{tian2018calibration}. In general, the calibration process involves measuring the phase noise of the local oscillator under different operating conditions and using the results to characterize the behavior of the system.

Instead of the phase fluctuations in the coherent state, the fluctuations in the amplitude quadrature are more versatile and feasible in the experimental context and do not require any interferometer, single-photon detectors, and complex calibration procedures. That the scheme is scalable  and can be used to generate large quantities of random numbers in real-time. In this experimental investigation, we have focused on this technique of the amplitude quadrature measurement through balanced homodyne detection and provided a complete overview to generate and certify the random bits generated through the technique.

\section{Theoretical Foundations of Entropy and Randomness}
Entropy is generally defined as the unavailability of energy or the measure of disorder in a system. Boltzmann was a pioneer in initiating the concept and presented a quantitative assessment of entropy the beautiful formula;
\eq{S=k\ln W,}
where $S$ stands for entropy, $k$ is the Boltzmann constant and $W=\frac{N!}{n!(N-n)!}$ represents the number of microstates that align with the given macrostate. Now, if we be capable of tracking every particle within the system, the entropy would reduce to zero. Here one notes the deep connection prevailing between entropy and the information, hitherto thought to be a subjective theme. later, in 1948, Shannon showed that less probable events convey more information than certain events. Mathematically,
\eq{{I(x)=\log(\frac{1}{P(x)})=-\log P(x)}\label{entrop_p}}
where $I(x)$ is the amount of information conveyed by the event $x$ an occurrence probability designated by $P(x)$. keeping Eq. \ref{entrop_p} in view, Shannon entropy can then be defined as a weighted average of the probabilities of all bits in the data string as below;
\eq{{H = -\sum_i p_i\log p_i,}\label{shan}}
where $P_i$ denotes the probability of the $i^{th}$ bit in a data string. From the above treatment, it is evident that higher probabilities of bits in a string result in less surprise i.e., less entropy, and vice versa. Hence entropy source as the best operational test to guess, patterns and degree of predictability prevailing in any data string or in the other words, entropy can efficiently evaluate the level of the randomness in any data set, whatsoever \cite{ben2008entropy,ben2008farewell}.
\subsection{Min Entropy}\label{Min entropy}
Min entropy is a concept in information theory that measures the unpredictability of a random variable. It represents the smallest amount of entropy that can be achieved by a given distribution and is calculated as the negative logarithm of the maximum probability in the distribution.  Mathematically, the min-entropy of a random variable $x$ is defined as $H_{\infty}(x) = -log(max{P(x)})$, where $P(x)$ is the probability of the observing outcome $x$. It is important to note that the min-entropy is always less than or equal to the Shannon entropy represented in Eq. \ref{shan} for any random variable $x$ and is typically measured in bits. For instance, consider the data string $'ABABCABA'$. The frequency of each symbol is $A=4, B=3, C=1$. The corresponding probability for symbols is $A=4/8=0.5, B=0.375, C=0.125$. The maximum probability is $0.5$, so the min-entropy of the string is $H_{\infty}(x) = -log(0.5) = 1$ bit.

In the context of QRNG, the min-entropy is used to characterize the randomness of the generated sequence of bits to ensure the security of cryptographic systems that rely on randomness. Essentially, the higher the min-entropy of a system, the more difficult it is for an attacker to guess or predict the information it contains. If the min-entropy of the QRNG sequence is too low, an adversary may be able to predict the random numbers with a significantly high probability.
\section{measuring the amplitude quadrature of quantum vacuum state: Theoretical Background}
\begin{figure}[htb]
	\includegraphics[scale=0.29]{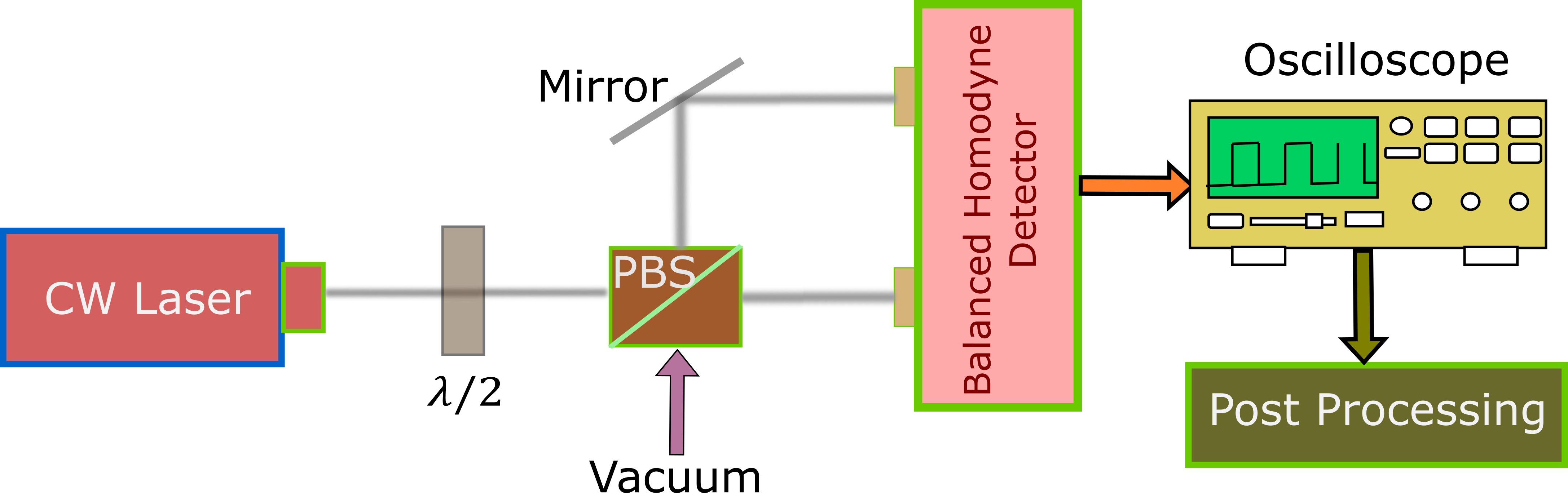}
	\caption{Experimental Schematics for QRNG based on Quantum Vacuum Noise in the amplitude quadrature of a photonic coherent state.}\label{fig:2atom}
\end{figure}
The schematic diagram for QRNG based on quantum vacuum noise is shown in Fig. \ref{fig:2atom}. A Continues Wave (CW) laser (Toptica) source of $852nm$, having an ultra-narrow linewidth of $50$kHz and a power tenability ranging from ($0-50$) mW is acting as a Local Oscillator (LO) followed by a $\lambda/2$ Half-Wave Plate (HWP) along with a Polarizing Beam Splitter (PBS). The intensity of the transmitted and the reflected parts of the beam after PBS is kept exactly equivalent by tuning the beam’s plane of polarization through HWP.The empty input port of PBS (lower) can be considered to be embedded with a quantum vacuum state that superimposes over the state of the local oscillator. A Balanced Homodyne Detector (BHD) comprised of two strictly identical photodiodes receives the output from both the arms and subtracts the photocurrents electronically. As the result, we are left only with the vacuum state noise accompanied by the noise of electronic components and that of the environment. The careful analysis of the entropy enables us to limit all the classical and electronic sources of the noise and to extract pure quantum noise in amplitude quadrature that follows Gaussian statistics. In the post-processing, the biased profile of quantum noise (Gaussian) is flattened by using Linear Feedback Shift Register (LFSR) as a potential extractor algorithm.

The photonic coherent state can be represented in a phase space of two quadratures (say $A_1$: (amplitude) X-axis and $A_2$: (phase) Y-axis) with equal uncertainties i.e., $\Delta A_1=\Delta A_2=1/2$. This state can be recognized on the phasor diagram as
\begin{align}
\kappa=A_1+\iota A_2=|\kappa| e^{\iota\phi},
\end{align}
with $\phi$ being the phase of the state and squared magnitude of the complex number $\kappa$ gives the average photon number $\bar{n}$ i.e., $|\kappa|^2=A_1^2+A_2^2=\bar{n}$. It can be noted that the quantum vacuum state corresponds to zero photons i.e., $|\kappa|^2=0$ and the fluctuations in both quadratures ($\Delta A_1$  and $\Delta A_2$) shift to the origin of the $2$D phasor plane and termed as vacuum fluctuations. The coherent state of LO and the vacuum state are given as
\begin{align}
\kappa_{lo}=\bar{\kappa}_{lo}+\delta A_{1,lo}(t)+\iota \delta A_{2,lo}(t),
\end{align}
and
\begin{align}
\kappa_{vac}=\delta A_{1,lo}(t)+\iota \delta A_{2,lo}(t),
\end{align}
respectively, where  $|\kappa_{lo} |^2$ is regarded as the mean photon number in the laser beam from LO. Mathematically, the transmitted and reflected beams from PBS have mixed field states as $w_1=(\kappa_{lo}+\kappa_{vac})/\sqrt2$ and $w_2=(\kappa_{lo}-\kappa_{vac})/\sqrt2$ respectively.
The photodiodes used in BHD are equally sensitive with the sensitivity $S$ when illuminated for the equal time interval $\tau$. They generate the photocurrents $i_1=S\hbar \omega\tau/w_1^2$ and $i_2=S\hbar \omega\tau/w_2^2$ which are then subtracted and amplified to obtain $\Delta i=i_2-i_1$. However, this difference of currents can be derived through brief mathematical manipulations and neglecting the small second-order terms (i.e., $\delta X^2$) as 
\eq{\Delta i=i_1-i_2=\frac{S\hbar\omega}{\tau}(2\bar\alpha_{lo}) \delta X_{1,vac}\label{deli}}
The above relation implies that the vacuum fluctuations ($\delta X_{1,vac}$) are directly relatable to the current difference and can thus be realized by using a BHD having true bandwidth.
\subsection{Balanced Homodyne Detection }\label{sec:TPB}
The difference of photocurrents $\Delta i$ in Eq. \ref{deli} can experimentally be realized by using a Balanced Homodyne Detector. It has been used extensively to detect the field quadrature of electromagnetic mode with the advancement of quantum technologies \cite{Zhang2021QuantumRN}. The true bandwidth of BHD, compatible with the shot noise along with the classical noise is one of the promising challenges against the detection of quantum to classical noise ratio. We designed a $10$MHz BHD using the two identical photodiodes (BPX $65$)
\begin{figure}[htb]
	\includegraphics[scale=0.8]{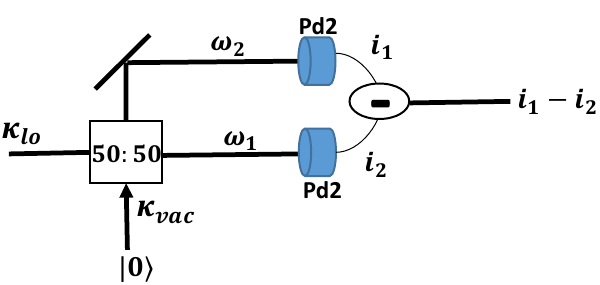}
	\caption{Schematic diagram of Balanced Homodyne Detector
	}\label{fig:2Bhd}
\end{figure}
having strictly similar characteristics. In Fig. \ref{fig:2Bhd}, the schematic diagram of BHD is shown. Here the state of the local oscillator ($\kappa_{lo}$) enters into one of the $50:50$ beam splitter ports whose other input port is inevitably embedded with vacuum state $\ket{0}$. The photodetectors Pd$1$ and Pd$2$ receive the coherent state of LO modulated with the quantum vacuum state $\omega_1$ and $\omega_2 $, respectively. The photocurrents generated through both the detectors are subtracted carefully to extract the vacuum noise or uncertainty in the amplitude quadrature of the vacuum state ($\delta X_{1,vac}$). The Eq. \ref{deli} depicts that the vacuum noise solely contributes to the current difference $\Delta i$ along with the inevitable electronic noise.
\section{Experimental Investigations, Results and Discussion}\label{sec:TPB}
\begin{figure}[htb]
	\includegraphics[scale=0.34]{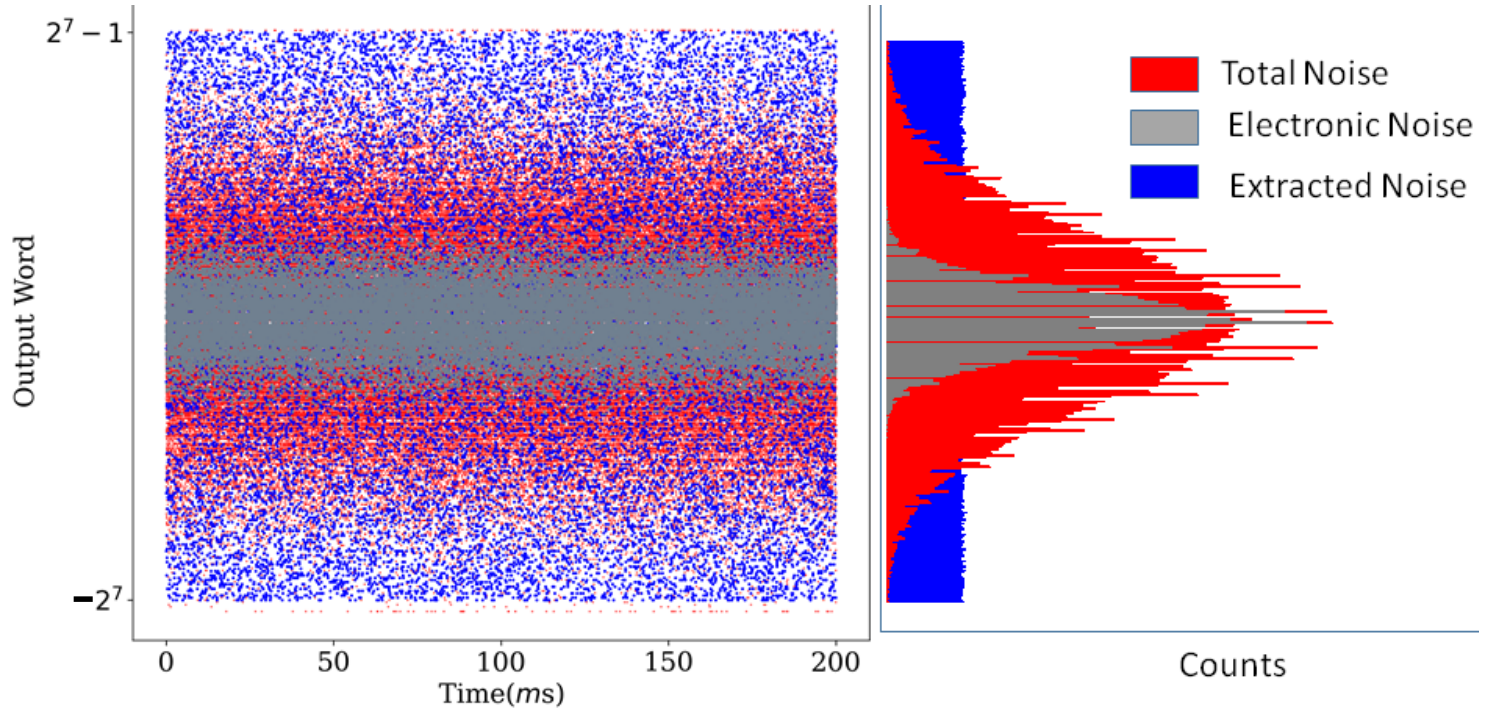}
	\caption{(left)Typical 1 million raw data collected under optimal sampling rate at 10MSa/s; (right) Histogram of the vacuum (red), electronic noise (gray) and random numbers extracted from the raw data based on LFSR-hashing extractor (blue)}\label{fig:2Noise}
\end{figure}
Using an $8$ bit sample, we marked the $2^8-1$ bins of identical sizes. Each bin is distinctly designated with an $8$-bit combination and the voltage counts corresponding to $\Delta i$ are obtained and marked into these bins randomly. Furthermore, these bins are symmetrically distributed around $0$ i.e., $-2^7<0<2^7-1$. In Fig. \ref{fig:2Noise} (left), the output words (counts) are sampled through Analog-to-Digital Converter (ADC) and plotted i.e., visualized on the oscilloscope against time delay. We quantify the electronic noise (gray) and total noise (red) in the absence and the presence of LO, respectively. The right panel of Fig. \ref{fig:2Noise} illustrates that both the noise follow the Gaussian profile with different variances. The Gaussian-like behavior is deterministic and introduces the biasing in the random numbers and is avoided by employing a $63$-bit LFSR extractor in serial on the raw bits (red). The extracted blue counts displayed in Fig. \ref{fig:2Noise} are equally probable in all bins and are maximally unpredictable.
The entropy evaluation of the random data bits characterizes the randomness and, for instance, the contribution of quantum noise in the raw data. We have used a sample comprising of $1$ Mega points to calculate worst-case min-entropy $H_{min}$ conditioned on the sampling rate of $10$ Mega samples/sec. The value of $H_{min}$ comes out to be about $5$, predicting that out of $8$ bit sample, the contribution of quantum noise is $5$ bits.
\begin{figure}[htb]
	\includegraphics[scale=0.71]{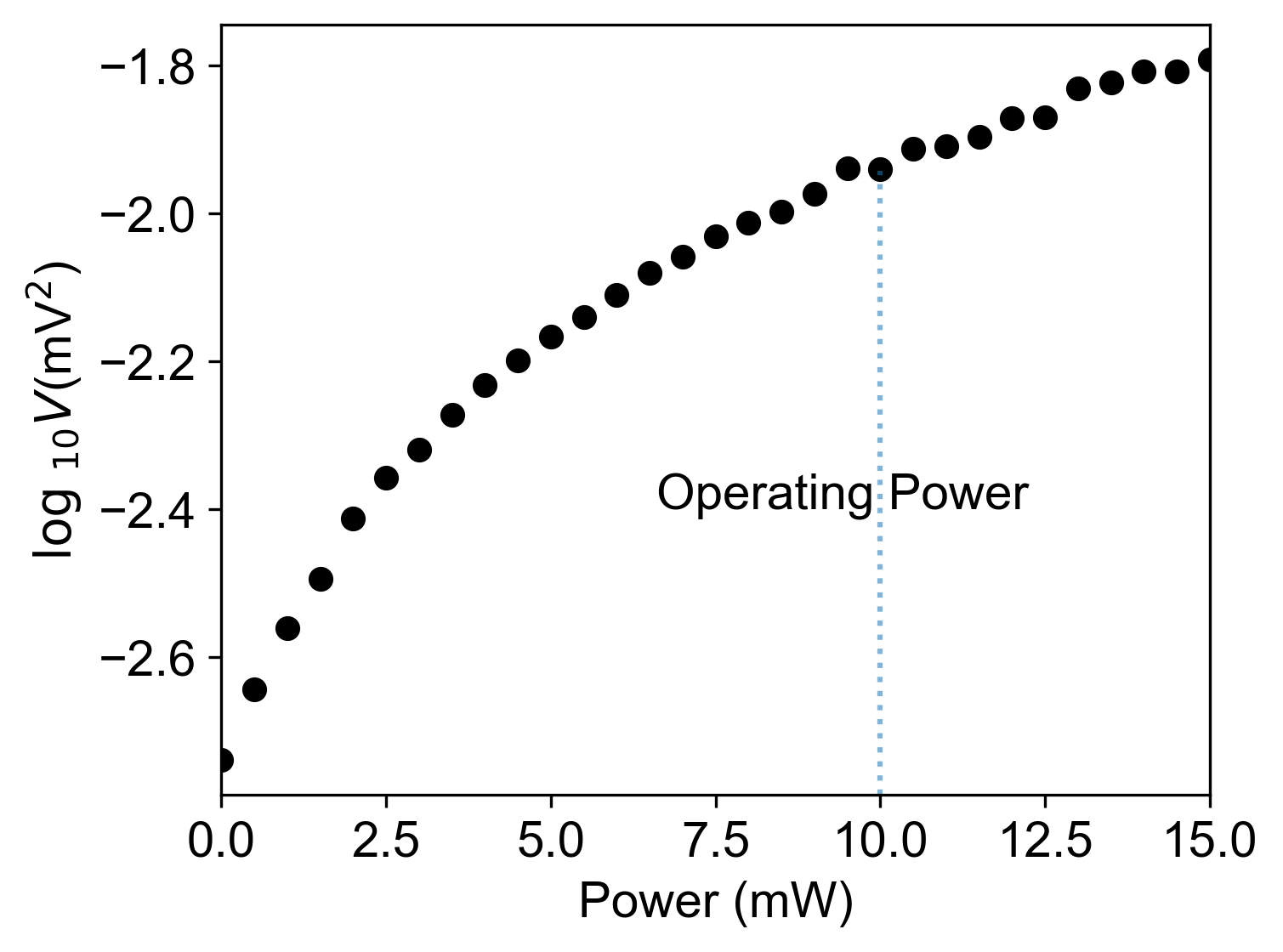}
	\caption{Evaluation and Characterizing the optimal  operating power of laser to maximize the variance of Gaussian noise profile. }\label{pv_temp}
\end{figure}
The variance of the Gaussian noise in the homodyne measurement depends on the power of the local oscillator \cite{gerry2005introductory}. At low powers, the variance increases with power as shown in Fig. \ref{pv_temp}. This is because the measurement is less precise due to the presence of shot noise, which is a fundamental limit to the precision of measurements. However, as the power of the local oscillator is increased, the variance eventually saturates and becomes roughly constant \cite{lei20208,zheng20196,guo2018enhancing}. This occurs because the shot noise is suppressed and the measurement becomes limited by other noise sources, such as the electronic noise in the detector. The point at which the variance starts to saturate or become constant is known as the "shot-noise limit" or the "quantum noise limit" \cite{gerry2005introductory}. Thus in order to obtain the maximum contribution of shot noise, we operate the laser at the power of $10$mW, the point of shot noise limit. In the case of a homodyne measurement, the Heisenberg uncertainty principle sets a fundamental limit on the precision of the measurement. It's important to note that the shot-noise limit is not the ultimate limit to the precision of a homodyne measurement. There are other sources of noise, such as phase noise and detector noise, that can limit the precision of the measurement even further.
\begin{figure}[htb]
	\includegraphics[scale=0.27]{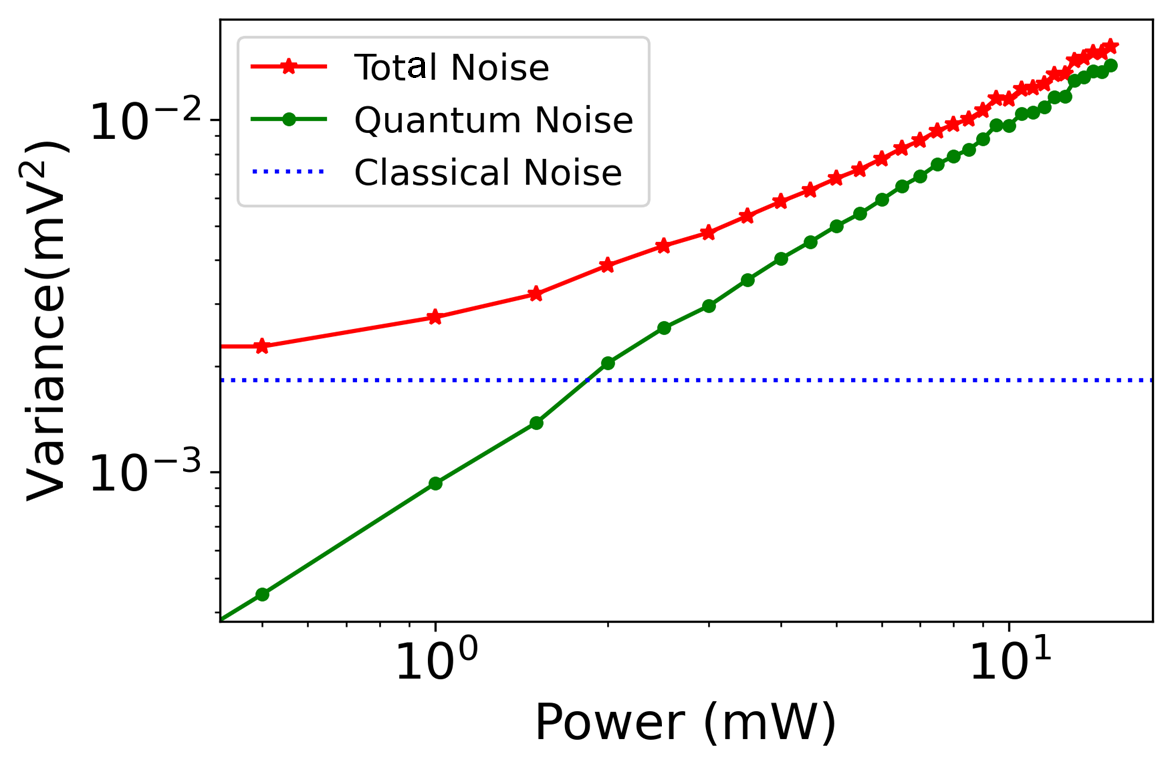}
	\caption{The logarithmic plot of variance of signal amplitude measured by the ADC as a function of LO equivalent power.}\label{fig:3PV}
\end{figure}
The power of LO (Laser) significantly influences the variance of Gaussian random noise. Apart from the quantum shot noise in homodyne detection, the major contribution comes from the intrinsic electronic noise in photodetectors as well as through the amplification circuitry. The noise in photodetectors includes thermal noise and dark current noise, both of which approximately follows Gaussian statistics. The variance of the noise against the different experimental values of LO power is plotted on logarithmic scales in Fig. \ref{fig:3PV}. The variance attributed to the total noise is designated as $\sigma_{total}$ and is comprised of two components i.e., $\sigma_{total}^2=\sigma_c^2+\sigma_q^2$ with $\sigma_c^2$ and $\sigma_q^2$ being the contribution from classical and quantum noise respectively. We measure the variance of classical noise ($\sigma_c^2$) by switching off the LO which remains fixed as all the classical factors including environmental noise, electronic components noise, etc., do not change with time throughout the experiment. However, slowly increasing the power of LO significantly increases the total variance ($\sigma_{total}^2$) and this increment solely comes from the quantum noise ($\sigma_q^2$). The blue dotted line in Fig. \ref{fig:3PV} is the classical noise which, as mentioned, is independent of LO power. However, variances of quantum (green) and total (red) noise increase with the offset of classical noise and coincide at the laser power of about $10$mW.
\subsection{NIST Entropy  Evaluation Suite (SP800-90B)}\label{sec:sp800}
 SP800-90B is a publication by the National Institute of Standards and Technology (NIST) that provides guidelines for assessing the entropy of random number generators (RNGs) along with an open-source software developed by NIST \cite{NISTSP800-90B}. The goal of the entropy assessment, as mentioned earlier, is to estimate the amount of entropy that is available in the RNG output. Further, as already stated, Entropy refers to the amount of randomness contained in the data, and it is measured in bits. The higher the entropy of the RNG output, the more secure the generated keys and data will be.

 The entropy estimators provided in SP800-90B are based on a statistical model of the RNG output data. This model assumes that the RNG output data is a sequence of independent and identically distributed (i.i.d.) random variables with a certain probability distribution. The entropy of the RNG output data is then estimated based on the probability distribution of the i.i.d. random variables.

There are two types of entropy estimators provided in SP800-90B: the min-entropy estimator and the collision entropy estimator. The min-entropy estimator provides an estimate of the minimum amount of entropy present in the RNG output, while the collision entropy estimator provides an estimate of the maximum amount of entropy present in the RNG output. The min-entropy estimator is based on the assumption that the probability distribution of the i.i.d. random variables is uniform. The entropy of the RNG output is then estimated as the negative logarithm of the maximum probability of any output sequence that could have been generated by the RNG. This estimator provides a conservative estimate of the entropy, as it assumes that the probability distribution is uniform and the worst-case scenario is considered.
By using the associated software to SP800-90B by NIST, we obtain $5$bits min-entropy per $8$bits of the sample that actually defines the source of randomness i.e., the quantum vacuum fluctuations. The estimated entropy is sufficient for credible random sources as reported earlier \cite{Chow2019CharacterizingRQ,ElBouazzati2020ImprovingRF,Borrelli2021EnhancingRO,Hilbig2019ImprovingRR}.
\subsection{Certification of Quantum Random Numbers}\label{sec:TPB}
 \textbf{Autocorrelation:}
\begin{figure}[htb]
	\includegraphics[scale=0.55]{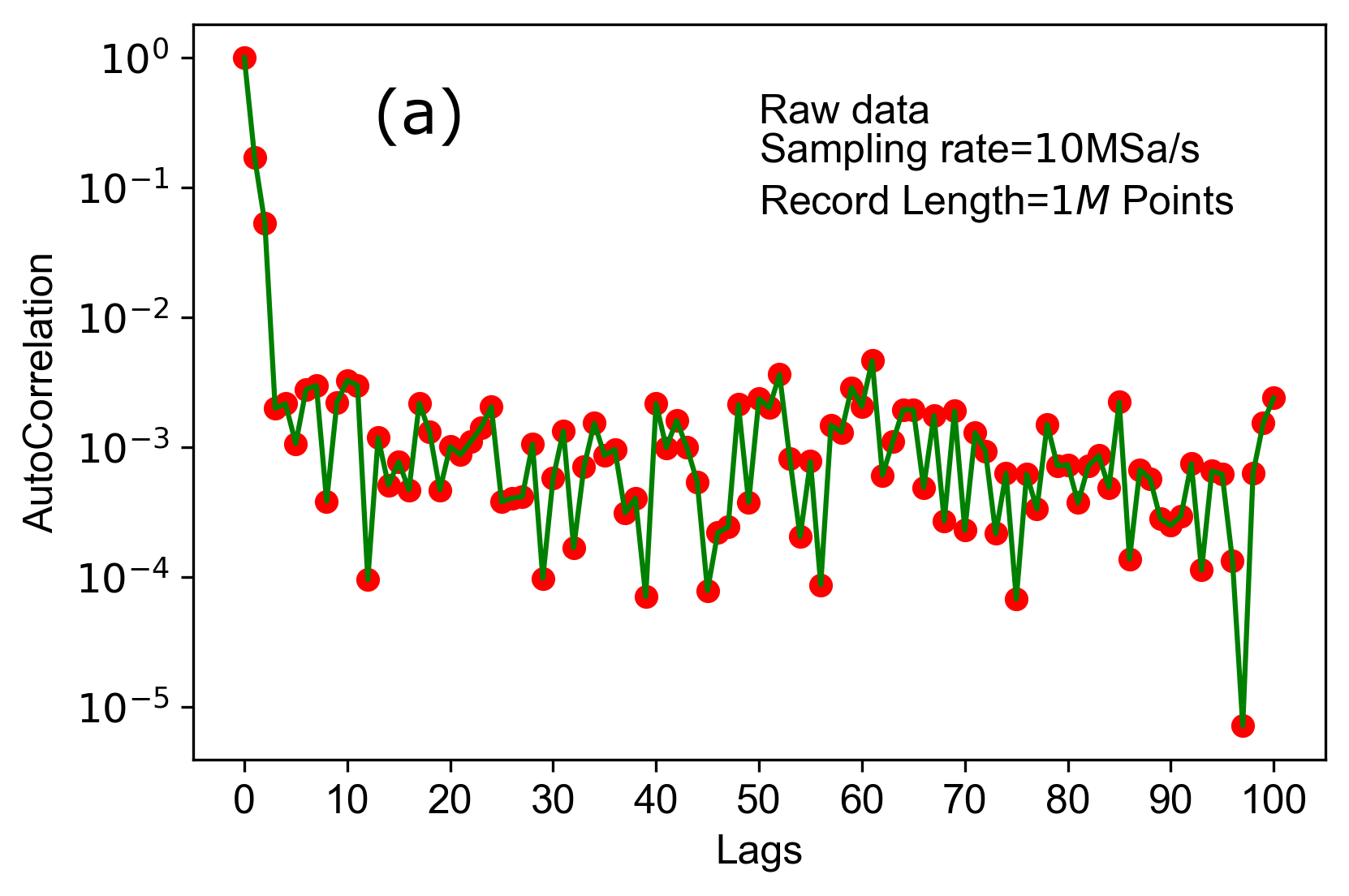}
	\includegraphics[scale=0.55]{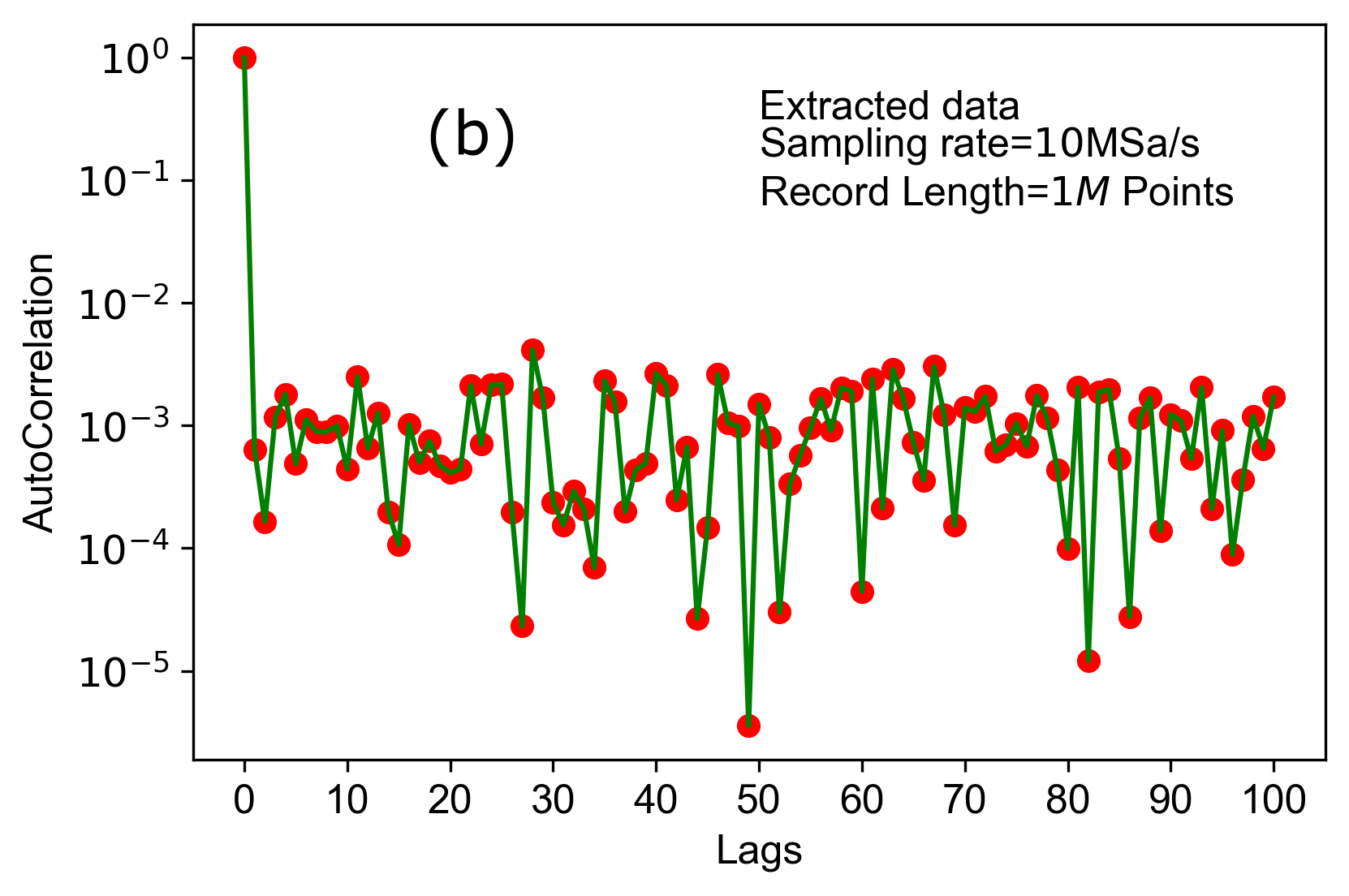}
	\caption{(a) Autocorrelation coefficient calculation of raw random data. The data size of each file is $10^6$ samples. (b) Autocorrelations coefficient calculation of the final random data after extraction. The data size of each file is $10^6$ bits}\label{fig:4acf}
\end{figure}
Now, in order to be sure that the extracted data out of the raw data exhibits random features, we need to perform certain mathematical transformations. Initially, the normalized autocorrelation coefficient $A(d)$,  defined as;
\begin{align}
A(d)=\frac{\langle x_i x_{i+d}\rangle_n}{\langle x_i^2 \rangle _n}
\end{align}
is characterized by $x_i$ and $x_(i+d)$ being the $8$ bits samples taken by the ADC in raw data with the sampling (temporal) delay $d$ of record length $n=10^6$ samples. It is important to note that the two samples are identical for no time delay ($d=0$) and thus be maximally correlated i.e., $A(0)=1$. The raw data is being sampled by $8$ bits ADC and we have discarded $3$ bits per sample (of $8$ bits) as the contribution of quantum noise is $5$ bits per sample suggested by min-entropy value i.e., $H_min=5$bits. Keeping the $5$ MSB bits and applying the LFSR in serial we obtain the extracted data. In Fig. \ref{fig:4acf}, raw and extracted data  are compared based on the autocorrelation coefficient $A(d)$. A significant correlation is found as shown in Fig. \ref{fig:4acf} (a) for the several sample lags as the two samples overlap temporally due to the short sampling rate. In the post-processing, it can be eradicated by flatting the Gaussian profile of the raw data through the LFSR extractor as shown in Fig. \ref{fig:4acf} (b). The output of the extractor is converted into bits and the lag is thus introduced in form of bits instead of samples in the extracted data. The autocorrelation coefficient has significantly been reduced to a few thousandths of the order soon after a lag of one bit as demonstrated by the data pattern in Fig. \ref{fig:4acf} (b). The residual value of autocorrelation can never drops to zero due to the finite bandwidth of the homodyne detector, certain modulating frequencies in electronic components as well as the other practical limitations.
\\ \textbf{Implementation of Compression Algorithms:}
\begin{figure}[htb]
	\includegraphics[scale=0.41]{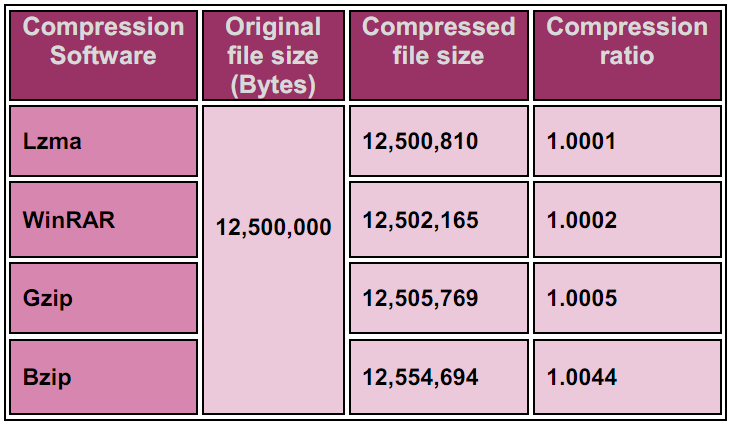}
	\caption{Comparison of original and compressed sizes of quantum random number file using different algorithms}\label{soft}
\end{figure}
While there are several tests to assess the randomness of random number generators, including the NIST statistical test suite and the Dieharder test suite, the compression test is an efficient way to obtain a quick estimate of the randomness of the random number file. The various compression algorithms including WinRAR, LZMA, Gzip, and Bzip were implemented. These algorithms are designed to identify and eliminate patterns in binary files. If the random number file is genuinely random, it should be difficult to compress. Fig. \ref{soft} shows the original file size, the final size after compression, and the ratio of the two sizes for each algorithm. A ratio close to $1$ indicates that the file cannot be efficiently compressed, indicating that it contains a high degree of randomness. Various studies have employed compression tests to evaluate the randomness of different random number generators, including quantum random number generators \cite{Stipcevic2013,Wang2016}. These studies have demonstrated that genuinely random data cannot be effectively compressed and that compression ratios close to $1$ are a promising indication of randomness.
\\ \textbf{NIST Test Suit (sts-2.1.2):}
\begin{figure}[htb]
	\includegraphics[scale=0.6]{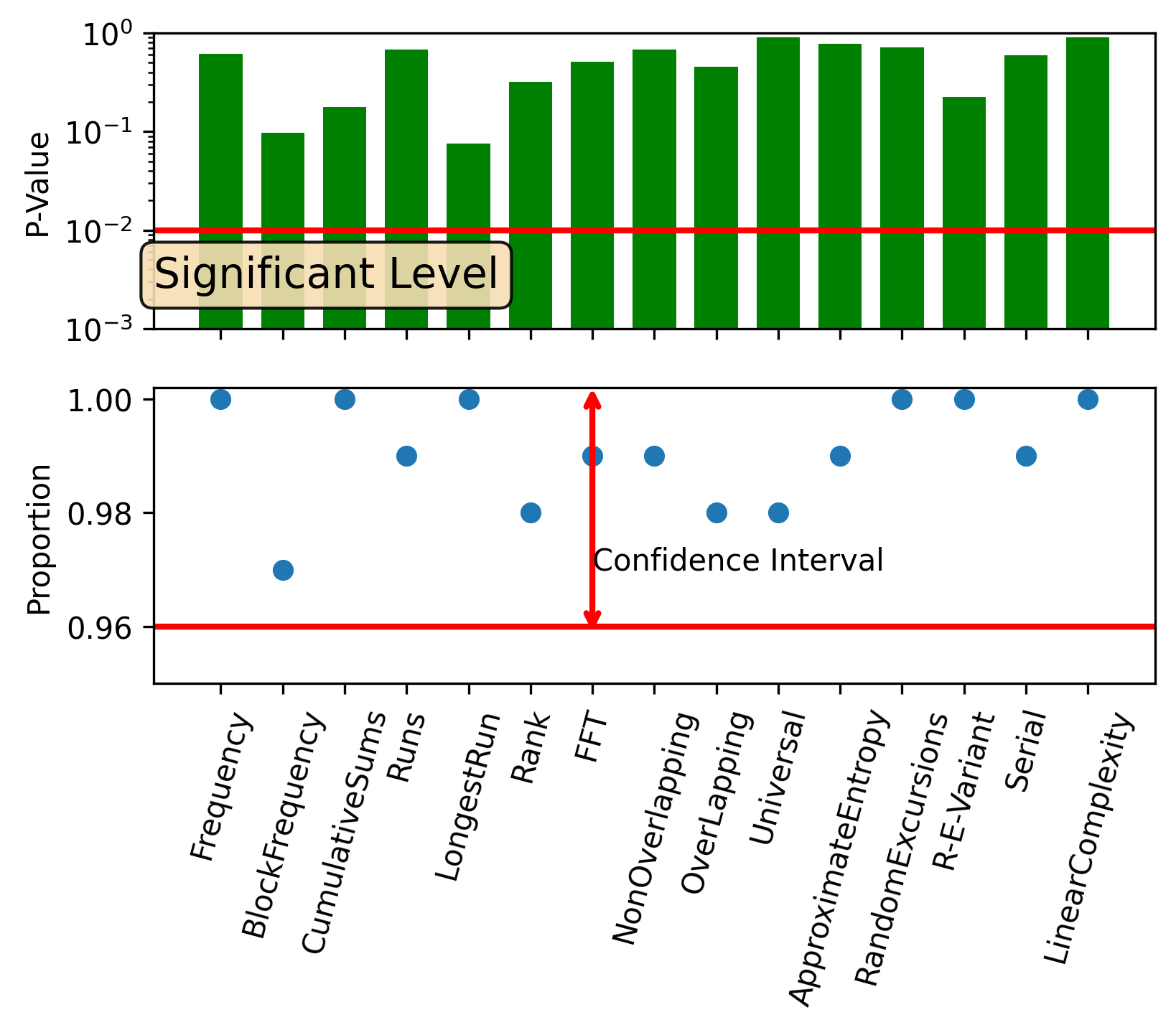}
	\caption{Typical results of standard NIST Statistical Test Suite. For each test term, the bar values represent the P values of the worst cases of our test outcomes. The red dots indicate that the minimum pass rate for each statistical test is $96\%$, which certifies a good level of true randomness}\label{fig:5N}
\end{figure}
The extracted random numbers were subjected to the National Institute of Standards and Technology (NIST) statistical test suite \cite{NISTSP80022}, which consists of $15$ rigorous statistical tests designed to evaluate the randomness and unpredictability of random number sequences. The final binary file of random numbers, which consisted of $100$ megabits ($12.5$ megabytes), was used for the NIST tests. To ensure the reliability and accuracy of the NIST test results, the sample size used for the tests was set to $0.1$ gigabits. The optimal parameters and a significant level for the P-value of $0.01$ were used, which is conventionally acceptable in relevant studies \cite{Wang2016}. The graphical report of all the NIST tests is shown in Fig. \ref{fig:5N}. Each test generates a P-value, which represents the probability of observing the test results by chance alone. A test is said to be successful if its corresponding P-value exceeds the significant level $\alpha$, marked by the horizontal in Fig. \ref{fig:5N}. In this study, all $15$ tests successfully passed the criterion, indicating that the random numbers generated through the presented technique exhibit a high degree of unpredictability. To further validate the unpredictability of the random numbers, the proportion of all the tests within the confidence interval of $(1 - \alpha)\pm 3p (1 - \alpha)\alpha/n$ was calculated and plotted on the shared x-axis of Fig. \ref{fig:5N} (lower). The confidence interval indicates the range within which the true proportion of successful tests lies with a certain level of confidence. The plot shows that all the tests fall within the confidence interval, providing additional evidence of the truly unpredictable nature of the random numbers.\\
\\ \textbf{Dieharder tests battery:}
\begin{figure}[htb]
	\includegraphics[scale=0.35]{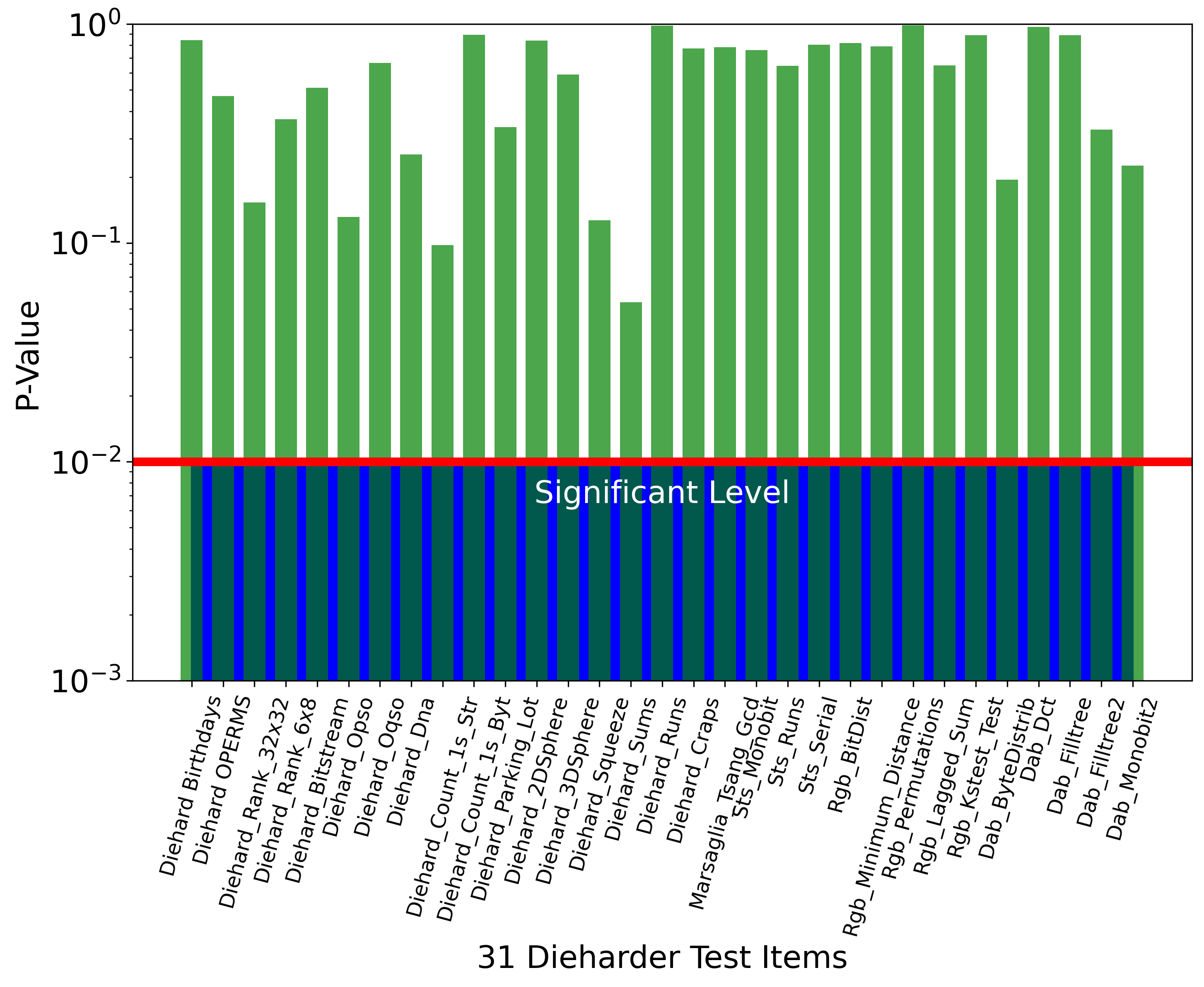}
	\caption{Dieharder test results for a sequence of obtained random numbers.}\label{die}
\end{figure}
The dieharder \cite{brown2018dieharder} battery of statistical tests is a highly reputed test suite and is extensively used to testify to the randomness of the data. It is a collection of over $31$ different statistical tests that are designed to evaluate the randomness of a given sequence of random numbers. These tests range from simple frequency and serial correlation tests to more advanced tests that use the properties of the Fourier transform and linear complexity to detect deviations from ideal randomness.
The test suite is highly configurable and can be used to test a wide range of RNG algorithms, including linear congruential generators, Mersenne Twister, and cryptographic algorithms such as AES and SHA-1. Fig. \ref{die}  displays the results of the Dieharder tests used to assess the randomness and provides the statistical significance of each test. The horizontal red line represents the significance level (alpha) of $0.01$, below which p-values are considered to be statistically significant. The green bars represent the p-values of each of the $31$ Dieharder tests, and the bars below the red line indicate tests where the p-value is not statistically significant. The y-axis is presented on a logarithmic scale, and the x-axis displays the name of each Dieharder test. A very large amount of data is required to fully characterize all of its tests (a random data file of $400$ Gigabits (GB) is used in ref. \cite{shi2016random}. However, the quality of our experimental setup and standard of the obtained random bits is so fine that all Dieharder tests qualify the passing criterion successfully on the data file of $50$GB. 
\section{Conclusion}
In this study, we have followed an operational approach to present the results of our experimental QRNG setup based on the amplitude quadrature measurement of the vacuum state. Of course, the entropy source of randomness i.e., photonic vacuum state elections is comparable to the Planck scale of $\hbar$. However, Local oscillators (LO) amplify them up to the extractable scale as demonstrated in Sec. V. From homodyne detection of LO states to the implementation of the LFSR extractor and digitization of random bits, the presented experimental setup is robust to the environment. Based on the finite tuning and precise adjustment through the half-wave plate, our setup becomes entirely source independent and leaves no depictable signatures of the source location and that of the specifically selected parameters. Further, our QRNG is highly efficient with a speed of about $50$ Mbps, that can easily be enhanced up to $500$ Mbps with a little alternation of the electronic circuitry and data manipulation gadgets. It can be accomplished by increasing the sampling rate of ADC limited by the bandwidth and sampling rate of BHD. Furthermore, bypassing the oscilloscope and designing the data acquisition system using the PCI card instead of USB to transfer the raw data significantly enhances the aggregate speed of the random numbers. However, increasing the data acquisition speed through ADC introduces more biasing in the raw data that requires rigorous hashing and extracting algorithms like the Teoplitz function in the step of post-processing, something that can be done with further elaborations. Another critical merit of our indigenously developed setup is that the schematics are not, in any way, affected by the photonic detection efficiency syndrome. The results of diverse characterizing measures including various reputed statistical test suites (NIST, Dieharder, etc.), autocorrelation coefficient, and NIST entropy assessment suite (SP800-90B) show the versatility of the random numbers. We believe that this guide may provide a sufficient resource of awareness to accomplish an experimental setup for QRNG on a laboratory scale as well as for a miniatured and handy device.
\section{Data Availability Statement}
As all experimental data has been presented in the main text graphically, therefore this manuscript has no associated data information.

\bibliography{qrng.bib}
\end{document}